\documentclass[%
reprint,
 amsmath,amssymb,
 aps,
 pre,
]{revtex4-2}

\usepackage{graphicx}% Include figure files
\usepackage{dcolumn}% Align table columns on decimal point
\usepackage{bm}% bold math
\usepackage{multirow}

\begin{document}

%\preprint{}

\title{Data-driven inference of Hopf normal form representations from oscillatory time series}


\author{Shinsuke Koyama}
%\email[]{skoyama@ism.ac.jp}
%\thanks{}
\affiliation{Department of Interdisciplinary Statistical Mathematics, The Institute of Statistical Mathematics, Tokyo 190-8562, Japan}

\author{Ryota Kobayashi}
%\email[]{}
%\thanks{}
\affiliation{Graduate School of Frontier Sciences, The University of Tokyo, Chiba 277-8561, Japan}
\affiliation{Mathematics and Informatics Center, The University of Tokyo, Tokyo 113-8656, Japan}


\date{\today}

\begin{abstract}
We introduce a data-driven framework that maps noisy oscillatory time series directly onto the Hopf normal form, enabling inference of underlying dynamics without knowledge of governing equations. 
By embedding the normal form in a probabilistic state-space model, the method jointly infers latent states and system parameters, yielding robust estimates of the natural frequency, Floquet exponent, and asymptotic phase even far from the bifurcation point and under strong noise. Combined with complex Gaussian process regression, the approach further reconstructs phase and amplitude sensitivity functions from data. Benchmarks on the van der Pol oscillator demonstrate substantially improved accuracy and noise robustness compared with existing phase-based and regression methods. This work establishes a direct bridge between normal-form theory and statistical inference, providing a general and practical route to low-dimensional descriptions of oscillatory dynamics in complex systems.   
\end{abstract}

%\keywords{}

\maketitle


%%%%%%%%%%%%%%%%%%%%%%%%%%%%%%%%%%%%%%%%%%%%
\section{Introduction}
%%%%%%%%%%%%%%%%%%%%%%%%%%%%%%%%%%%%%%%%%%%%

Rhythmic activity is a ubiquitous phenomenon observed across a wide range of biological systems. Representative examples include oscillatory dynamics in cortical networks of the brain \citep{Buzsaki04,Wang10}, circadian rhythms in mammals \citep{Mohawk11}, cardiac and respiratory cycles in humans \citep{Schafer98,Kralemann13,Kobayashi16}, and locomotor patterns such as animal gait \citep{Collins93,Kiehn06,Funato16}. Owing to their repetitive and self-sustained nature, such phenomena are commonly modeled as limit cycle oscillators \citep{Strogatz18,Winfree01}, which provide a fundamental mathematical framework for describing stable periodic behavior. Studying limit cycle oscillators is essential for understanding the dynamics of oscillatory systems, as well as for enabling the inference and control of their behavior in both natural and artificial settings.  

A canonical mechanism underlying the emergence of limit cycle oscillations is the Hopf bifurcation. In this scenario, a stable fixed point characterized by a pair of complex-conjugate eigenvalues loses stability as a system parameter varies, leading to the birth of a stable limit cycle from infinitesimal perturbations. Because of its structural robustness and universality, the Hopf bifurcation serves as a central theoretical paradigm for explaining the onset of rhythmic behavior in diverse physical and biological systems. 

To analyze the dynamics near bifurcation points in a unified framework, nonlinear dynamical systems can often be simplified through an appropriate change of variables into a normal form \citep{Guckenheimer83,Wiggins03,Kuznetsov04}. The normal form framework provides a systematic way to study dynamics near bifurcation points by reducing complex behavior to a tractable low-dimensional system, while preserving the qualitative properties of the original system through topological equivalence. In particular, systems undergoing a Hopf bifurcation are locally topologically equivalent to a cubic normal form, which captures the essential mechanism responsible for the emergence and stabilization of a limit cycle. 

The normal form framework has advanced our theoretical understanding of nonlinear dynamics \citep{Hoppensteadt97}. However, its application to the quantitative analysis of observed data remains relatively limited. While a coordinate transformation to the normal form is mathematically guaranteed to exist near bifurcation points, the explicit functional form of this transformation is generally unknown. As a result, identifying normal form representations directly from data remains a challenging problem. Recently, a data-driven framework has been proposed to infer dynamical systems near a Hopf bifurcation from estimated phase and amplitude response curves \citep{Wilson25}. Nevertheless, inferring representations consistent with the normal form from noisy observations remains challenging, particularly in the presence of substantial observation noise.

The primary objective of this article is to bridge normal form theory and data-driven analysis.  Specifically, we propose a method for inferring representations consistent with the Hopf normal form from noisy observations of oscillatory activity governed by unknown dynamical equations. This representation enables the extraction of intrinsic dynamical properties of the underlying system in a data-driven manner. We achieve this by embedding the Hopf normal form within a state-space modeling framework \citep{Durbin01,Kitagawa10}, which enables the inference of coordinates consistent with the normal form. 

The proposed framework consists of two components. The first component is based on a state-space approach that estimates model parameters, including the natural frequency and the nontrivial Floquet exponent, from noisy oscillatory time-series data. This approach additionally assigns a latent state, representing phase and amplitude, to each observation. The second component extends these pointwise estimates by interpolating the complex-valued latent states over unobserved regions of the observation space using complex-valued Gaussian process regression. In particular, this approach enables the estimation of phase- and amplitude-sensitivity functions. Compared with existing approaches, the proposed framework exhibits improved robustness to observation noise while accurately inferring the underlying dynamical properties.   

The main contributions of this study are summarized as follows:   
\begin{enumerate}   
  \item    We propose a data-driven method for constructing representations consistent with the Hopf normal form from noisy observational time-series data, without requiring prior knowledge of the governing equations.     
  \item     We develop a state-space--based model for robust estimation of system parameters and latent phase--amplitude states under observation noise.
  \item     We infer intrinsic dynamical properties, including phase and amplitude sensitivity functions, from the inferred representations. 
\end{enumerate}

The remainder of this article is organized as follows. 
Section~\ref{sec:hopf} briefly reviews the Hopf normal form. Section~\ref{sec:method} presents the proposed methodology, with detailed derivations and algorithmic procedures provided in the Appendix. Section~\ref{sec:results} demonstrates the performance of the proposed approach using synthetic data generated from a van der Pol oscillator. Finally, Section~\ref{sec:discussion} concludes the article with a summary of the main findings and a discussion of potential extensions and implications.

%%%%%%%%%%%%%%%%%%%%%%%%%%%%%%%%%%%%%%%%%%%%
\section{Hopf normal form}
\label{sec:hopf}
%%%%%%%%%%%%%%%%%%%%%%%%%%%%%%%%%%%%%%%%%%%%
In this section, we briefly review the Hopf normal form, which serves as the foundation for the proposed method developed in the subsequent sections. Consider an autonomous dynamical system described by the vector field
\begin{equation}
\frac{d\bm{y}(t)}{dt} = F(\bm{y}(t)),
\quad \bm{y}(t) \in \mathbb{R}^D,
\quad (D \ge 2),
\label{eq:ode_hopf}
\end{equation}
and assume that the systesm possesses a stable periodic orbit arising from a fixed point via a supercritical Hopf bifurcation. In a neighborhood of the Hopf bifurcation point, the system can be transformed, via an appropriate smooth change of variables, into a canonical normal form known as the Stuart--Landau equation \citep{Kuramoto84,Nakao16}. For the system \eqref{eq:ode_hopf}, we consider the following canonical form in the vicinity of the Hopf bifurcation point:
\begin{equation}
\frac{d}{dt}
\begin{pmatrix}
x_1 \\
x_2
\end{pmatrix}
=
\begin{pmatrix}
\alpha x_1 - \omega x_2 - \alpha x_1 \left(x_1^2 + x_2^2\right) \\
\omega x_1 + \alpha x_2 - \alpha x_2 \left(x_1^2 + x_2^2\right)
\end{pmatrix},
\label{eq:normal_form}
\end{equation}
where $\omega > 0$ denotes the angular frequency and $\alpha > 0$ determines the rate of attraction toward the limit cycle. 
Introducing polar coordinates $(r,\theta)$, the system \eqref{eq:normal_form} can be rewritten as
\begin{align}
\frac{d\theta}{dt} &= \omega, \label{eq:hopf_w} \\
\frac{dr}{dt} &= \alpha(r - r^3). \label{eq:hopf_r}
\end{align}
These equations show that the system admits a stable periodic orbit on the unit circle, along which the motion is a uniform rotation with constant angular velocity $\omega$. 
The parameter $\alpha$ governs the exponential convergence of trajectories toward the limit cycle and is directly related to the Floquet exponent of the limit-cycle by $\lambda = -2\alpha$. 
According to Hopf bifurcation normal form theory, the original system \eqref{eq:ode_hopf} is locally topologically equivalent to \eqref{eq:normal_form} in a neighborhood of the bifurcation point, ensuring that the qualitative dynamical behavior of the system is preserved under the transformation. 
In Appendix~\ref{appendix:phase_amplitude}, we review the conventional phase--amplitude representation of limit-cycle oscillators \citep{Kuramoto84,Wedgwood13,Nakao16,Shirasaka17,shirasaka2020phase} and discuss its relationship to the Hopf normal form.

%%%%%%%%%%%%%%%%%%%%%%%%%%%%%%%%%%%%%%%%%%%%
\section{Data-driven inference of Hopf normal form representations}   \label{sec:method}
%%%%%%%%%%%%%%%%%%%%%%%%%%%%%%%%%%%%%%%%%%%%

The objective of this study is as follows. Given oscillatory time-series data generated by a dynamical system with unknown governing equations, we seek to infer a representation from the observations that is consistent with the Hopf normal form \eqref{eq:normal_form}. This transformation enables principled, data-driven extraction of dynamical characteristics of the underlying process.

\begin{figure*}
\centering
\includegraphics[width=0.95\textwidth]{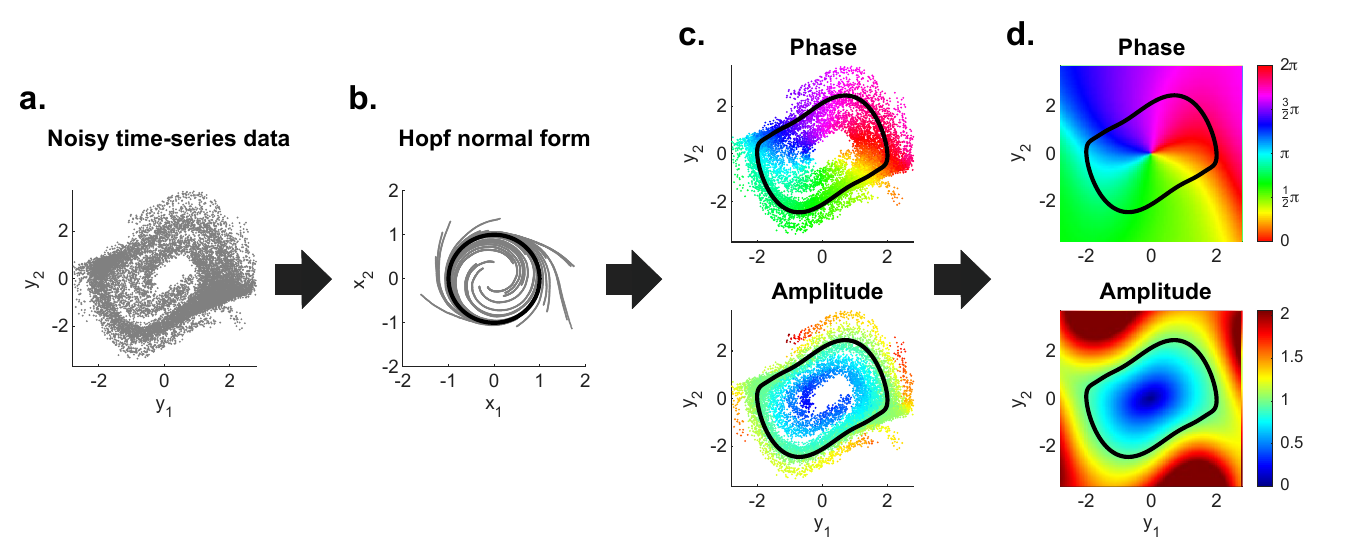}
\caption{
Schematic illustration of the proposed method.
(a) Noisy oscillatory time-series data observed in the measurement space.
(b) Latent dynamics represented in the Hopf normal-form coordinates, inferred from the observations using the state-space model.
(c) Pointwise assignment of the estimated phase and amplitude to each observation in the data space.
(d) Interpolation of the latent phase and amplitude over unobserved regions of the observation space using complex Gaussian process regression (CGPR).
}
\label{fig:schem_method}
\end{figure*}

Figure~\ref{fig:schem_method} provides a schematic overview of the proposed framework. In the first stage, we estimate the model parameters, including the natural frequency $\omega$ and the nontrivial Floquet exponent $\lambda$, from noisy time-series data using a state-space approach. Based on the estimated parameters, the state-space model further infers the latent states associated with the Hopf normal form (Figure~\ref{fig:schem_method} a, b). In the second stage, these pointwise latent-state estimates are extended over unobserved regions of the observation space by interpolating the complex-valued states using complex Gaussian process regression (CGPR) (Figure~\ref{fig:schem_method} c, d). 
The first stage of the proposed method is presented in Subsections~\ref{sec:ssm}, \ref{sec:state_inf}, and \ref{sec:param_est}, which describe the state-space formulation, latent-state inference, and parameter estimation, respectively.  
The second stage is presented in Subsection~\ref{sec:Rec_Phase_Amp}, which introduces the interpolation of the latent states based on complex Gaussian process regression.

%%%%%%%%%%%%%%%%%%%%
\subsection{State space model}		\label{sec:ssm}
We consider discrete-time measurements obtained with sampling period $\Delta$. Henceforth, $t = 1,2,\ldots$ denotes discrete time. Let $\bm{y}_t \in \mathbb{R}^D$ ($D\ge 1$) be the observed vector at time $t$, measured from oscillatory activity.
For each observation $\bm{y}_t$, we introduce a latent state vector $\bm{x}_t \in \mathbb{R}^2$ that evolves according to the discrete-time counterpart of the normal form \eqref{eq:normal_form}. 
Using a first-order Euler discretization of the normal form \eqref{eq:normal_form} with additive noise,  the state dynamics are given by
\begin{align}
\bm{x}_{t+1} = 
\begin{pmatrix}
1+\alpha\Delta & -\omega\Delta \\ 
\omega\Delta & 1+\alpha\Delta
\end{pmatrix}
\bm{x}_t - \alpha\Delta|\bm{x}_t|^2\bm{x}_t + \bm{\xi}_t,
\label{eq:state_model}
\end{align}
where $\bm{\xi}_t \sim \mathcal{N}(\bm{0}, \Sigma_x)$ is zero-mean Gaussian white noise with the covariance matrix $\Sigma_x$. The inclusion of process noise enables us to embed the normal form within a state-space framework and accounts for modeling inaccuracies and state estimation errors in practice.

We assume that the observed vector is generated from the latent state through a nonlinear mapping $\bm{h} : \mathbb{R}^2 \to \mathbb{R}^D$:
\begin{equation}
\bm{y}_t = \bm{h}(\bm{x}_t) + \bm{\zeta}_t,
\label{eq:observation_model}
\end{equation}
where $\bm{\zeta}_t \sim \mathcal{N}(\bm{0}, \Sigma_y)$ denotes Gaussian observation noise.
The mapping $\bm{h}$ plays a central role in linking the observed variables to the latent normal form coordinates. We model it using multivariate polynomials of degree $p$:
\begin{align}
\bm{h}(\bm{x})=
C_0 + C_1 \bm{x} +C_2 \bm{x}^{(2)} + \cdots + C_p \bm{x}^{(p)},
\label{eq:polynomial}
\end{align}
where 
$\bm{x}^{(p)} = (x_1^p, x_1^{p-1}x_2, x_1^{p-2}x_2^2, \ldots, x_2^p )^\mathsf{T} \in \mathbb{R}^{p+1}$  is a vector
and 
$C_p \in \mathbb{R}^{D \times (p+1)}$ is a matrix. 

The state equation \eqref{eq:state_model} together with the observation model \eqref{eq:observation_model} defines a nonlinear state-space model. For an observed dataset $\bm{y}_{1:T} \equiv \{\bm{y}_1,\ldots,\bm{y}_T\}$, the likelihood is given by
\begin{eqnarray}
p(\bm{y}_{1:T} \mid \Theta) =&& 
\int d\bm{x}_{0:T} p(\bm{x}_0 \mid \Theta) \nonumber\\
&&\times
\prod_{t=1}^Tp(\bm{x}_t \mid \bm{x}_{t-1},\Theta)p(\bm{y}_t \mid \bm{x}_t,\Theta),
\label{eq:likelihood}
\end{eqnarray}
where 
$\Theta = \{ \alpha, \omega, \Sigma_x, C_{0:p}, \Sigma_y, \bm{\mu}_0, \Sigma_0\}$ 
denotes the collection of model parameters, and the initial distribution is
$p(\bm{x}_0 \mid \Theta) = \mathcal{N}(\bm{\mu}_0,\Sigma_0)$.

%%%%%%%%%%%%%%%%%%%%%%%%%%%%%%%%
\subsection{Latent state inference with Gaussian moment matching}
\label{sec:state_inf}
%%%%%%%%%%%%%%%%%%%%%%%%%%%%%%%%
We now describe how to infer the posterior distribution over the latent states $\bm{x}_{0:T}$ given observations $\bm{y}_{1:T}$ and parameters $\Theta$. As in standard state-space modeling, inference proceeds in two stages: forward filtering to compute $p(\bm{x}_t \mid \bm{y}_{1:t}, \Theta)$, followed by backward smoothing to obtain $p(\bm{x}_t \mid \bm{y}_{1:T}, \Theta)$.
Since both the state transition and observation models are nonlinear, exact inference is intractable. We therefore employ a Gaussian approximation based on moment matching \citep{Deisenroth09,Turner10,Donner25}.

%%%%%%%%%%%%%%%%%%%%%%%%
\subsubsection{Forward iteration (Filtering)}
%%%%%%%%%%%%%%%%%%%%%%%%
Suppose that at time $t-1$ the filtering distribution is Gaussian:
$p(\bm{x}_{t-1}|\bm{y}_{1:t-1},\Theta) = \mathcal{N}(\bm{\mu}^f_{t-1},\Sigma^f_{t-1})$.
Our goal is to compute $p(\bm{x}_t \mid \bm{y}_{1:t}, \Theta)$ from $p(\bm{x}_{t-1}|\bm{y}_{1:t-1},\Theta)$.
We first evaluate the predictive density:
\begin{eqnarray}
p(\bm{x}_{t} \mid \bm{y}_{1:t-1},\Theta) =&&
 \int d\bm{x}_{t-1} p(\bm{x}_{t} \mid \bm{x}_{t-1},\Theta) \nonumber\\
&&\times  
p(\bm{x}_{t-1} \mid \bm{y}_{1:t-1},\Theta). 
\label{eq:prediction}
\end{eqnarray}
Due to the nonlinearity of the state equation \eqref{eq:state_model}, this integral cannot be computed analytically. We therefore approximate the predictive density by a Gaussian distribution whose mean and covariance match the exact first and second moments:
\begin{align}
p(\bm{x}_{t} \mid \bm{y}_{1:t-1},\Theta)  
\approx \mathcal{N}(\bm{\mu}^p_t,\Sigma^p_t),
\end{align}
where $\bm{\mu}_t^p$ and $\Sigma_t^p$ are obtained via moment matching (see Appendix~\ref{appendix:filtersmoothing}).

Using this Gaussian approximation, the filtering density is computed via Bayes' rule:
\begin{align}
p(\bm{x}_{t} \mid \bm{y}_{1:t},\Theta) 
&= \frac{p(\bm{y}_t \mid \bm{x}_t,\Theta) p(\bm{x}_t \mid \bm{y}_{1:t-1},\Theta)}
{p(\bm{y}_t \mid \bm{y}_{1:t-1},\Theta)  } \nonumber\\
&\approx
\mathcal{N}(\bm{\mu}_t^f,\Sigma_t^f). 
\label{eq:filtering}
\end{align}
The explicit expressions for $\bm{\mu}_t^f$ and $\Sigma_t^f$ are provided in Appendix~\ref{appendix:filtersmoothing}. This completes the forward pass.

%%%%%%%%%%%%%%%%%%%%%%%%
\subsubsection{Backward iteration (Smoothing)}
%%%%%%%%%%%%%%%%%%%%%%%%
Next, we compute the smoothing distribution $p(\bm{x}_t|\bm{y}_{1:T},\Theta)$. 
Starting from the final filtering distribution at $t=T$: $p(\bm{x}_{T}|\bm{y}_{1:T},\Theta)$, we proceed backward in time using
\begin{eqnarray}
p(\bm{x}_{t} \mid \bm{y}_{1:T},\Theta) =&&
\int d\bm{x}_{t+1} p(\bm{x}_{t} \mid \bm{x}_{t+1}, \bm{y}_{1:t},\Theta) \nonumber\\
&&\times 
p(\bm{x}_{t+1} \mid \bm{y}_{1:T},\Theta).
\label{eq:smoothing}
\end{eqnarray}
The conditional density is given by
\begin{equation}
p(\bm{x}_{t} \mid \bm{x}_{t+1}, \bm{y}_{1:t},\Theta) = 
\frac{p(\bm{x}_{t+1} \mid \bm{x}_t,\Theta)p(\bm{x}_t \mid \bm{y}_{1:t},\Theta)}
{p(\bm{x}_{t+1} \mid \bm{y}_{1:t},\Theta)}.
\end{equation}
Again, because of the nonlinearity in \eqref{eq:state_model}, exact evaluation is intractable. We therefore apply Gaussian moment matching, yielding the approximation
\begin{equation}
p(\bm{x}_t|\bm{y}_{1:T},\Theta) \approx \mathcal{N}(\bm{\mu}^s_t,\Sigma^s_t).
\end{equation}
Details of the smoothing updates are given in Appendix~\ref{appendix:filtersmoothing}.

%%%%%%%%%%%%%%%%%%%%%%%%%%%%%%%%
\subsection{Parameter estimation}		\label{sec:param_est}
%%%%%%%%%%%%%%%%%%%%%%%%%%%%%%%%
In the previous section, we described approximate inference over the latent states given fixed parameters $\Theta$. We now turn to parameter estimation.
We adopt the Expectation--Maximization (EM) algorithm \citep{Dempster77,Shumway82} to estimate $\Theta$. Rather than maximizing the likelihood \eqref{eq:likelihood} directly, we iteratively maximize the auxiliary function
\begin{eqnarray}
\mathcal{Q}(\Theta,\Theta') =&&
\mathbb{E}_{\Theta}[\log p(\bm{x}_0 \mid \Theta')] \nonumber\\
&&+ 
\sum_{t=1}^T\{ \mathbb{E}_{\Theta}[\log p(\bm{y}_t \mid \bm{x}_t,\Theta')] \nonumber\\
&&+ 
\mathbb{E}_{\Theta}[\log p(\bm{t}_t \mid \bm{x}_{t-1},\Theta')]
\},
\label{eq:qfunc}
\end{eqnarray}
where the expectations are taken with respect to the smoothing distributions obtained under the current parameter values $\Theta$.
Given current estimates $\Theta$, the updated parameters are obtained by maximizing $\mathcal{Q}(\Theta, \Theta')$ with respect to $\Theta'$. The E- and M-steps are iterated until convergence. Explicit update formulas are provided in Appendix~\ref{appendix:em}.

%%%%%%%%%%%%%%%%%%%%%%%%%%%%%%%%
\subsection{Reconstruction of phase and amplitude sensitivity functions}		\label{sec:Rec_Phase_Amp}
%%%%%%%%%%%%%%%%%%%%%%%%%%%%%%%%
As described above, the first stage of the proposed method estimates the model parameters $\Theta$ from the observed time-series data $\{\bm{y}_t\}$ using the EM algorithm. Given the estimated parameters $\hat{\Theta}$, the latent states $\{\bm{\mu}_t^s := \mathbb{E}_{\Theta}[ \bm{x}_{t} \mid \bm{y}_{1:T} ] \}$ are subsequently inferred via forward--backward smoothing. In the second stage, these pointwise latent-state estimates are extended over unobserved regions of the observation space to construct continuous phase and amplitude functions. To achieve this, we employ complex Gaussian process regression (CGPR), a nonparametric Bayesian framework for modeling complex-valued functions \citep{Tortosa18}.

First, the estimated latent state at each time point, $\bm{\mu}_t^s$, is transformed into the complex-valued representation $\hat{z}_t = \hat{r}_t e^{i\hat{\theta}_t}$, where $(\hat{r}_t,\hat{\theta}_t)$ denote the polar coordinates corresponding to $\bm{\mu}_t^s$. Using the paired observations and latent representations, $\{\bm{y}_t,\hat{z}_t\}$, as training data, we interpolate the complex-valued state at arbitrary locations in the observation space via CGPR. 

Given $N$ training samples $\{(\bm{y}_t,\hat{z}_t)\}_{t=1}^N$, CGPR provides the predictive mean and variance of the complex-valued state at an input point $\bm{y}_*$ as
\begin{align}
\hat{z}(\bm{y}_*) &= \bm{k}_*^\dagger (K + \sigma_z^2 I)^{-1} \hat{\bm{z}}, \label{eq:mean_CGPR} \\
v(\bm{y}_*) &= k(\bm{y}_*, \bm{y}_*) 
- \bm{k}_*^\dagger (K + \sigma_z^2 I)^{-1} \bm{k}_*, \label{eq:var_CGPR}
\end{align}
where $k(\bm{y}_i,\bm{y}_j)$ denotes the kernel function describing the correlation between outputs $\hat{z}_i$ and $\hat{z}_j$, $K_{ij}=k(\bm{y}_i,\bm{y}_j)$, $\bm{k}_* = (k(\bm{y}_1,\bm{y}_*),\ldots,k(\bm{y}_N,\bm{y}_*))^\mathsf{T}$, $\hat{\bm{z}}=(\hat{z}_1,\ldots,\hat{z}_N)^\mathsf{T}$, and $\sigma_z^2$ represents the variance of the proper complex Gaussian noise.

In this study, we adopt the Gaussian kernel
\begin{equation}
k(\bm{y}_i, \bm{y}_j) = 
a^2 \exp\left(-\frac{1}{2} (\bm{y}_i - \bm{y}_j)^\mathsf{T} \Sigma^{-1} (\bm{y}_i - \bm{y}_j)\right),
\end{equation}
where $\Sigma$ is a positive-definite diagonal matrix. The hyperparameters $\{\sigma_z^2, a^2, \Sigma\}$, which determine the noise level and kernel shape, are estimated by maximizing the log marginal likelihood
\begin{align}
L = -\hat{\bm{z}}^\dagger (K + \sigma_z^2 I)^{-1}\hat{\bm{z}} 
- \log \det (K + \sigma_z^2 I) - N \log \pi.
\end{align}

The predictive mean \eqref{eq:mean_CGPR} defines a continuous complex-valued function extended from the latent-state estimates $\{\bm{\mu}_t^s\}$. The phase $\hat{\theta}(\bm{y}_*)$ and amplitude $\hat{R}(\bm{y}_*) = 1-\hat{r}(\bm{y}_*)$ at an arbitrary point $\bm{y}_*$ in the observation space are obtained from the complex representation
\begin{equation}
\hat{z}(\bm{y}_*)
=
\hat{r}(\bm{y}_*)
\exp\!\left(i\hat{\theta}(\bm{y}_*)\right).
\end{equation}
Here, the amplitude $\hat{R}(\bm{y}_*)$ is defined so that it vanishes on the limit cycle. This definition differs from the conventional amplitude function \citep{Wedgwood13,Shirasaka17}. It is defined with respect to the coordinate system induced by the Hopf normal form \eqref{eq:normal_form}. Appendix~\ref{appendix:phase_amplitude} provides details of the conventional phase--amplitude representation of limit-cycle oscillators and its relationship to the Hopf normal form.

In particular, we apply this framework to estimate the phase sensitivity function (PSF), $\bm{Z}(\theta)$, and the amplitude sensitivity function (ASF), $\bm{I}(\theta)$. These functions are defined as the gradients of the phase and amplitude functions evaluated along the limit cycle and characterize the linear response properties of phase and amplitude, respectively. Further details on the PSF and ASF are provided in Appendix~\ref{appendix:phase_amplitude}. Using CGPR, these quantities are approximated by finite differences as
\begin{align}
\hat{Z}_j(\theta_i) = 
\frac{\hat{\theta}(\bm{y}^{lc}(\theta_i) + \epsilon \bm{e}_j) - \hat{\theta}(\bm{y}^{lc}(\theta_i) - \epsilon \bm{e}_j)}{2\epsilon},
\label{eq:ePSF}
\end{align}
and
\begin{align}
\hat{I}_j(\theta_i) = 
\frac{\hat{R}(\bm{y}^{lc}(\theta_i) + \epsilon \bm{e}_j) - \hat{R}(\bm{y}^{lc}(\theta_i) - \epsilon \bm{e}_j)}{2\epsilon},
\label{eq:eASF}
\end{align}
where $\bm{e}_j$ denotes the unit vector in the $j$th coordinate direction and $\epsilon$ is a small constant, set to $10^{-5}$ in this study. 
The limit-cycle trajectory in the observation space is obtained as 
\begin{equation}
\bm{y}^{lc}(\theta)
=
\bm{h}(\bm{x}^{lc}(\theta)),
\label{eq:limitcycle_est}
\end{equation}
where $\bm{x}^{lc}(\theta)=(\cos\theta,\sin\theta)$, $\theta \in [0,2\pi)$, represents the unit circle in the latent space.

%%%%%%%%%%%%%%%%%%%%%%%%%%%%%%%%%%%%%%%%%%%%
\section{Numerical experiments}\label{sec:results}
%%%%%%%%%%%%%%%%%%%%%%%%%%%%%%%%%%%%%%%%%%%%
We apply the proposed method to oscillatory time-series data to assess its ability to extract the dynamical properties of the underlying process.

%%%%%%%%%%%%%%%%%%%%%%
\subsection{Benchmark system: van der Pol oscillator}

We evaluate the proposed method using synthetic data generated from the van der Pol (vdP) oscillator, to assess its accuracy and robustness to observation noise. The vdP oscillator is given by
\begin{equation}
\frac{d}{dt}
\begin{pmatrix}
y_1 \\
y_2
\end{pmatrix} 
= 
\begin{pmatrix}
y_2 \\
\gamma(1-y_1^2)y_2 - y_1
\end{pmatrix},
\label{eq:vdp}
\end{equation}
where $\gamma$ is the bifurcation parameter. Equation~\eqref{eq:vdp} has a fixed point at $(y_1,y_2)=(0,0)$, which undergoes a supercritical Hopf bifurcation at $\gamma=0$. For $\gamma>0$, a stable limit cycle emerges around the fixed point. In what follows, we generate time-series data in this oscillatory regime ($\gamma>0$).

The vdP oscillator is simulated using the fourth-order Runge--Kutta method with time step $\Delta = 0.01$ over a total duration of $T = 1000$. Let $\{\bm{y}_t^\ast\}$ denote the underlying noiseless trajectory. We generate 100 trajectories from initial conditions sampled uniformly from the region $-2.5 \le y_1, y_2 \le 2.5$. Gaussian white noise with zero mean and standard deviation $\sigma_y$ is then added to the trajectories to obtain noisy observations, denoted by $\{\bm{y}_t\}$, to which the proposed state-space method is applied. 
Simulations are performed for varying values of the bifurcation parameter $\gamma \in \{ 0.2,  0.4,  0.6,  0.8,  1.0  \}$ and observation noise level $\sigma_y \in \{ 0.01, 0.1,  1.0  \}$ in order to investigate how the estimation accuracy deteriorates as the system moves farther from the bifurcation point and as the observation noise increases.

%%%%%%%%%%%%%%%%%%%%%%
\subsection{Estimation procedure}
We describe the procedure to estimate the system parameters, including the natural frequency, Floquet exponent, and the asymptotic phase and to reconstruct the phase sensitivity function (PSF) and amplitude sensitivity function (ASF) from time series.

%%%%%%%%%%%%%%%%%%%%%%%%
\subsubsection{Proposed method}
%%%%%%%%%%%%%%%%%%%%%%%%

The model parameters of the state-space model $\Theta$ are first estimated from the set of time series using the EM algorithm (see Subsection~\ref{sec:param_est} for details). Given the estimated parameters $\hat{\Theta}$, the latent states $\{\bm{\mu}_t^s\}$ for each time series are inferred by forward--backward smoothing (see Subsection~\ref{sec:state_inf} for details). Using the estimated polynomial coefficients $\hat{C}_0, \cdots, \hat{C}_p$ in \eqref{eq:polynomial}, the latent states are mapped to the observation space as
\begin{align}
\hat{\bm{y}}_t 
= \hat{\bm{h}}(\bm{\mu}_t^s) 
\equiv \hat{C}_0 + \hat{C}_1 \bm{\mu}^s_t + \hat{C}_2 (\bm{\mu}^s_t)^{(2)} + \cdots + \hat{C}_p (\bm{\mu}^s_t)^{(p)},
\label{eq:y_denoised}
\end{align}
which serves as a denoised estimate of the observations. In this study, the polynomial order in \eqref{eq:polynomial} is fixed at $p = 5$. In particular, the limit-cycle orbit is estimated using \eqref{eq:limitcycle_est}, where the observation function is replaced with $\hat{\bm{h}}$. The resulting estimated limit-cycle orbit is subsequently used in \eqref{eq:ePSF} and \eqref{eq:eASF} to estimate the PSF and ASF, respectively.

To apply CGPR for interpolating the estimated states to unobserved points in the observation space, we use the set of denoised observations and corresponding complex-valued states, $\{\hat{\bm{y}}_t, \hat{z}_t\}$, as training data. Denoised observations are employed instead of the raw observations in order to mitigate the effect of observation noise. To reduce the computational cost of CGPR, $N = 1000$ data points are uniformly sampled to construct the training set. The PSF and ASF are then estimated using \eqref{eq:ePSF} and \eqref{eq:eASF}.

%%%%%%%%%%%%%%%%%%%%%%%%
\subsubsection{Baseline methods}
%%%%%%%%%%%%%%%%%%%%%%%%

For comparison, we apply an empirical approach to estimate the frequency, Floquet exponent, and asymptotic phase from noisy time-series data \citep{Namura22,Yamamoto25}. First, a moving average with window width 0.07 is applied to reduce observation noise. The period is estimated as $\hat{T}_0 = (s_{n+1}-s_1)/n$, where $s_1$ and $s_{n+1}$ denote the first and $(n+1)$-th crossings of a Poincar\'e section, respectively. The frequency is then given by $\hat{\omega} = 2\pi / \hat{T}_0$.
Then, we estimate the Floquet exponent using the method of~\citet{Wolf85} for Lyapunov exponent estimation. For two-dimensional limit-cycle oscillators, the Floquet exponents coincide with the Lyapunov exponents. Since the exponent associated with the phase direction is zero, it suffices to estimate the sum $\lambda_1 + \lambda_2$ by tracking the temporal evolution of a small phase-space area.
Finally, the asymptotic phase is estimated using a direct method \citep{Yamamoto25}. For a time series $\{\bm{y}_t\}$ whose endpoint $\bm{y}_T$ lies sufficiently close to the limit cycle, the phase $\hat{\theta}_T$ is obtained via linear interpolation along the cycle. The phase at earlier times is then given by $\hat{\theta}_t = \hat{\theta}_T + \hat{\omega}(t-T)$.

For PSF and ASF estimation, we compare the proposed approach with two existing methods: the derivative phase regression (DPR)~\citep{Namura22}, which estimates both phase and amplitude functions based on polynomial regression, and Gaussian Process Phase Interpolation (GPPI)~\citep{Yamamoto25}, which estimates only the phase function using real-valued Gaussian process regression.

%%%%%%%%%%%%%%%%%%%%%%
\subsection{Performance metrics}

We evaluate the performance of the proposed method in terms of its ability to recover the dynamical properties of the underlying system. Specifically, we assess the extent to which the estimated natural frequency $\hat{\omega}$ and the nonzero Floquet exponent $\hat{\lambda} = -2\hat{\alpha}$ of the Hopf normal form approximate those of the vdP oscillator. 
In addition, denoising performance is quantified by the root mean square error (RMSE) between the ground-truth trajectories $\{\bm{y}_t^\ast\}$ and the denoised estimates $\{\hat{\bm{y}}_t\}$.

Next, we evaluate the accuracy of asymptotic phase estimation. The polar angle $\hat{\theta}_t$ of the estimated latent state $\bm{\mu}_t^s$ can be interpreted as the asymptotic phase. This is because the latent state rotates with approximately constant angular velocity $\hat{\omega}$ (see Appendix~\ref{appendix:phase_amplitude} for the definition of the asymptotic phase). We therefore compare $\hat{\theta}_t$ with the asymptotic phase $\theta_t$ of the vdP oscillator. By convention, the phase origin is defined as the point on the periodic orbit at which the first coordinate attains its maximum value. The similarity between the asymptotic phase $\theta_t$ and its estimate $\hat{\theta}_t$ is quantified by
\begin{equation}
	\rho = \overline{\cos(\theta_t - \hat{\theta}_t)},
\label{eq:phase_corr}
\end{equation}
where the overline denotes averaging over all time points. Values of $\rho$ closer to 1 indicate higher  estimation accuracy.

To evaluate the accuracy of the estimated PSF $\hat{\bm{Z}}(\theta)$ and ASF $\hat{\bm{I}}(\theta)$, we use the coefficient of determination~\citep{Namura22,Yamamoto25}:
\begin{align}
R^2_Y = 1 - 
\frac{\sum_i \big( Y_j(\theta_i) - \hat{Y}_j(\theta_i) \big)^2}
{\sum_i \big( Y_j(\theta_i) - \overline{Y}_j \big)^2},
\end{align}
where $Y_j$ denotes either $Z_j$ or $I_j$, and $\overline{Y}_j$ is the mean of $\{Y_j(\theta_i)\}$. Values of $R_Y^2$ closer to 1 indicate higher estimation accuracy. The ground-truth PSF $\hat{\bm{Z}}(\theta)$ and ASF $\hat{\bm{I}}(\theta)$ of the vdP oscillator are computed numerically using the adjoint method~\citep{Takata21,Namura22}.

%%%%%%%%%%%%%%%%%%%%%%
\subsection{Results} 

\begin{figure*}
\includegraphics[width=0.95\textwidth]{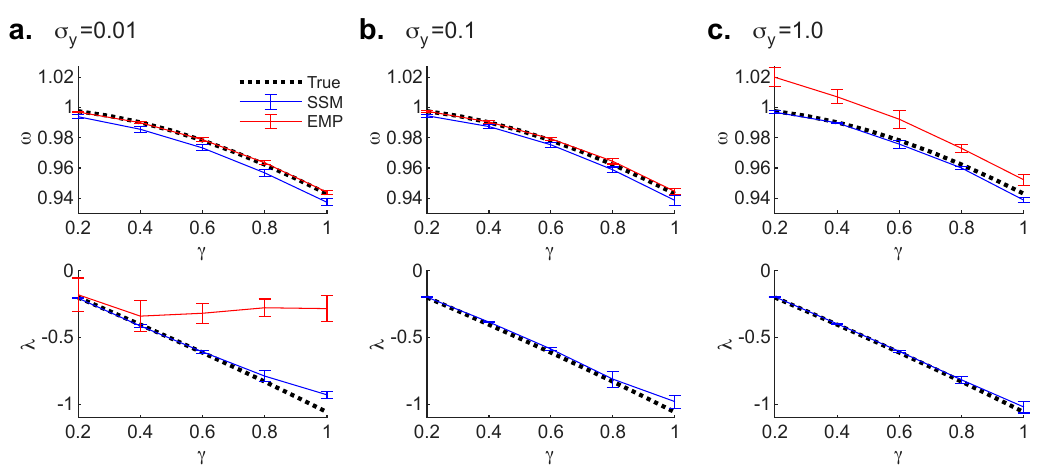}
\caption{
Estimation of natural frequency and Floquet exponent.   
Estimated natural frequency $\omega$ (top panels) and Floquet exponent $\lambda$ (bottom panels) as functions of the bifurcation parameter $\gamma$, obtained using the proposed state-space method (SSM, blue) and the empirical method (EMP, red). Error bars indicate the standard deviation across 10 independent realizations, and the dotted line denotes the ground-truth value. The observation noise levels are (a) $\sigma_y = 0.01$, (b) $\sigma_y = 0.1$, and (c) $\sigma_y = 1.0$. For the empirical method, estimates of the Floquet exponent at $\sigma_y = 0.1$ and $\sigma_y = 1.0$ are omitted in panels (b) and (c), respectively, because the estimated values fall outside the plotted range. 
}
\label{fig:freqfloq}
\end{figure*}

We present the results of numerical experiments on the vdP oscillator. First, we examine the estimation accuracy of the model parameters. Figure~\ref{fig:freqfloq} shows the estimated natural frequency $\hat{\omega}$ and nonzero Floquet exponent $\hat{\lambda}$ as functions of the bifurcation parameter $\gamma$, obtained using the proposed state-space method (SSM; blue), together with the ground-truth values (dotted black). Overall, the proposed method provides accurate estimates of these quantities over a wide range of $\gamma$ and $\sigma_y$. Notably, reliable estimation is achieved even far from the bifurcation point and under substantial observation noise.
For comparison, results obtained using the empirical method are also shown in the figure (EMP; red). Although the frequency can be estimated reasonably accurately, estimation of the Floquet exponent deteriorates even under moderate observation noise. These results demonstrate that the proposed method is substantially more robust to observation noise in estimating dynamical parameters.

\begin{figure}
\includegraphics[width=0.45\textwidth]{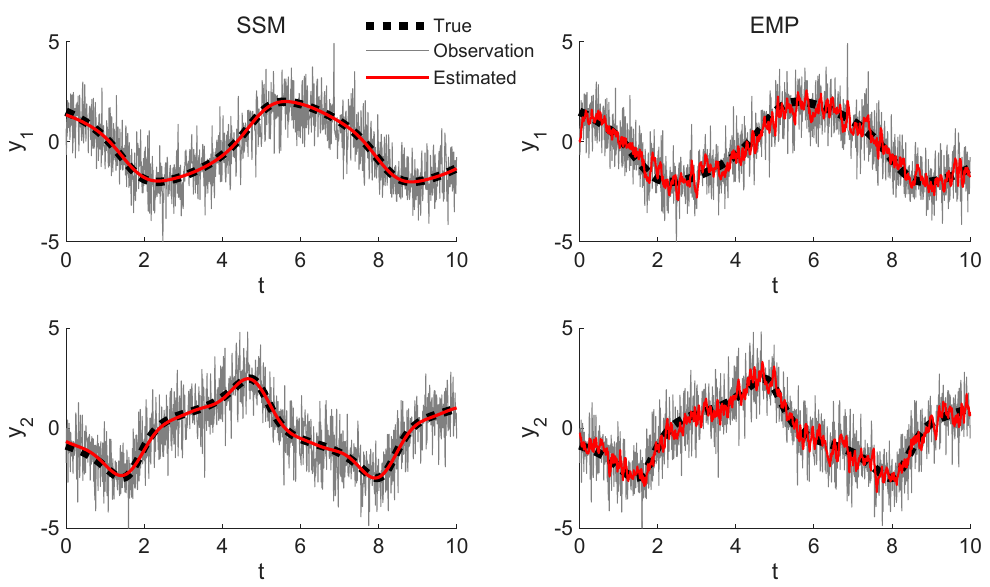}
\caption{
Denoising of oscillatory time series.
The black dotted, gray, and red curves denote the original signal $\bm{y}_t^*$, the observed noisy signal $\bm{y}_t$ corrupted by Gaussian white noise with $\sigma_y = 1.0$, and the estimated signal $\hat{\bm{y}}_t$, respectively. 
The left panel shows results obtained by the proposed state-space method (SSM), while the right panel shows those obtained by a moving-average approach (EMP).
}
\label{fig:denoise}
\end{figure}

To further evaluate robustness to noise, we compare the denoising performance of the proposed method with that of a moving average filter. Figure~\ref{fig:denoise} presents representative time series, including the original signal (dotted black), the noisy observations (gray), and the denoised estimates (red). The proposed method yields substantially smoother and more accurate reconstructions. Table~\ref{tb:rmse} reports the RMSE values, confirming that the proposed method consistently outperforms the moving average approach.

\begin{table}[h]
\caption{\label{tb:rmse}
Comparison of the root mean squared error (RMSE) between the true (noiseless) and denoised time series, generated by the proposed state-space method (SSM) and a moving-average empirical method (EMP). Methods with the lowest RMSE are highlighted in {\bf bold}.  
}
\begin{ruledtabular}
\begin{tabular}{ccccc}
 $\gamma$ & Method & $\sigma_y=0.01$ & $\sigma_y=0.1$ & $\sigma_y=1.0$ \\ \hline
\multirow{2}{*}{0.2} &  SSM & {\bf 0.003} & {\bf 0.011} & {\bf 0.052} \\
&EMP & 0.106 & 0.118 & 0.544\\ \hline
\multirow{2}{*}{0.4} &  SSM & {\bf 0.003} & {\bf 0.013} & {\bf 0.053}  \\
&EMP & 0.107 & 0.119 & 0.543  \\ \hline
\multirow{2}{*}{0.6} &  SSM & {\bf 0.004} & {\bf 0.017} & {\bf 0.089} \\
&EMP & 0.109 & 0.121 & 0.544  \\ \hline
\multirow{2}{*}{0.8} &  SSM & {\bf 0.004} & {\bf 0.021} & {\bf 0.088}  \\
&EMP & 0.111 & 0.123 & 0.545  \\ \hline
\multirow{2}{*}{1.0} &  SSM & {\bf 0.005} & {\bf 0.025} & {\bf 0.121}  \\
&EMP & 0.114 & 0.126 & 0.546 
\end{tabular}
\end{ruledtabular}
\end{table}

Next, we evaluate the estimation accuracy of the asymptotic phase using the metric defined in \eqref{eq:phase_corr}.  Table~\ref{tb:phase} summarizes the results for the proposed state-space method (SSM) and the empirical method (EMP). The proposed method consistently achieves high accuracy across all tested values of $\gamma$ and $\sigma_y$, whereas the performance of the empirical method deteriorates markedly as the observation noise increases.

\begin{table}[h]
\caption{\label{tb:phase}%
Comparison of the phase similarity metric $\rho$ between the proposed state-space method (SSM) and the empirical method (EMP). Methods with the highest $\rho$ are highlighted in {\bf bold}. 
}
\begin{ruledtabular}
\begin{tabular}{ccccc}
$\gamma$ & Method & $\sigma_y=0.01$ & $\sigma_y=0.1$ & $\sigma_y=1.0$ \\ \hline
\multirow{2}{*}{0.2} &  SSM & 0.998 & {\bf 0.999}  & {\bf 0.999} \\
&EMP & {\bf 0.999} & 0.996  & 0.817 \\ \hline
\multirow{2}{*}{0.4} &  SSM & 0.997 & 0.998 & {\bf 0.999}  \\
&EMP & {\bf 0.999} & 0.998 & 0.826  \\ \hline
\multirow{2}{*}{0.6} &  SSM & 0.997 & 0.998 & {\bf 0.998} \\
&EMP & {\bf 0.999} & 0.998 & 0.828  \\ \hline
\multirow{2}{*}{0.8} &  SSM & 0.997 & 0.998 & {\bf 0.999}  \\
&EMP & {\bf 0.999} & 0.998 & 0.818  \\ \hline
\multirow{2}{*}{1.0} &  SSM & 0.998 & {\bf 0.998} & {\bf 0.999}  \\
&EMP & {\bf 0.999} & 0.997 & 0.806
\end{tabular}
\end{ruledtabular}
\end{table}

\begin{figure*}
\centering
\includegraphics[width=0.95\textwidth]{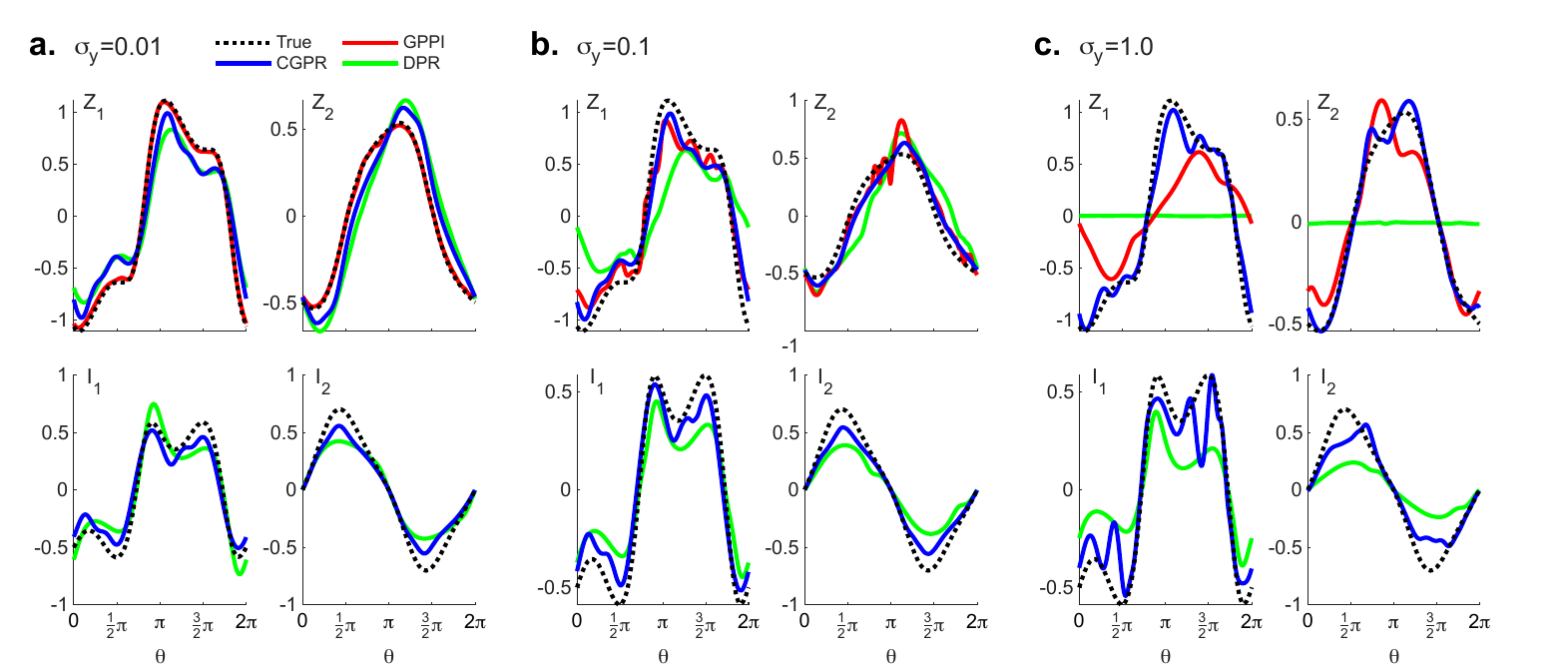}
\caption{
Reconstruction of phase and amplitude sensitivity functions. 
Phase sensitivity function (top panels) and amplitude sensitivity function (bottom panels) estimated by the proposed method (CGPR, blue) and existing methods (GPPI, red; DPR, green). The dotted curves indicate the ground truth for the van der Pol (vdP) oscillator with $\gamma = 0.8$. The observation noise levels are (a) $\sigma_y = 0.01$, (b) $\sigma_y = 0.1$, and (c) $\sigma_y = 1.0$. Note that GPPI is not plotted in the bottom panels, as it estimates only the phase sensitivity function.
}
\label{fig:PSF_ASF}
\end{figure*}

Finally, we assess the estimation accuracy of the phase sensitivity function (PSF) and amplitude sensitivity function (ASF) by comparing the proposed method with two existing methods---the derivative phase regression method (DPR)~\citep{Namura22} and Gaussian Process Phase Interpolation (GPPI)~\citep{Yamamoto25}. 
Note that GPPI estimates only the PSF. Figure~\ref{fig:PSF_ASF} presents representative examples of the estimated PSF and ASF together with the corresponding ground-truth functions. Tables~\ref{tb:PSF} and \ref{tb:ASF} summarize the average coefficients of determination for the PSF and ASF, respectively, computed over two coordinate directions and 10 independent realizations. Overall, the proposed method accurately estimates both the PSF and ASF across a broad range of parameter settings. Although GPPI achieves slightly higher PSF accuracy under low observation noise, its performance degrades as the noise level $\sigma_y$ increases. In contrast, the proposed method remains comparatively robust to observation noise owing to the denoised training data obtained from the state-space model. The performance of DPR in estimating both the PSF and ASF is comparable to that of the proposed method under low observation noise, but similarly deteriorates as the noise level increases.

\begin{table}[h]
\caption{\label{tb:PSF}%
Comparison of the coefficient of determination ($R^2_Z$) for the phase sensitivity function (PSF) between the proposed method (CGPR) and existing methods (DPR and GPPI). 
Methods with the highest $R^2_Z$ are highlighted in {\bf bold}.  
}
\begin{ruledtabular}
\begin{tabular}{ccccc}
$\gamma$ & Method & $\sigma_y=0.01$ & $\sigma_y=0.1$ & $\sigma_y=1.0$ \\ \hline
\multirow{3}{*}{0.2} &  CGPR & 0.951 & 0.942 & {\bf 0.980} \\
&DPR & 0.948 & 0.935 & 0.000 \\
&GPPI & {\bf 0.997} & {\bf 0.986} & 0.919 \\ \hline
\multirow{3}{*}{0.4} &  CGPR & 0.871 & 0.901 & {\bf 0.990}  \\
&DPR & 0.885 & 0.815 & 0.001 \\
&GPPI & {\bf 0.999} & {\bf 0.973} & 0.834  \\ \hline
\multirow{3}{*}{0.6} &  CGPR & 0.826 & 0.915 & {\bf 0.941} \\
&DPR & 0.846 & 0.705 & 0.001 \\
&GPPI & {\bf 0.999} & {\bf 0.969} & 0.757  \\ \hline
\multirow{3}{*}{0.8} &  CGPR & 0.848 & 0.918 & {\bf 0.964}  \\
&DPR & 0.821 & 0.629 & 0.001 \\
&GPPI &  {\bf 0.998} & {\bf 0.931} & 0.703  \\ \hline
\multirow{3}{*}{1.0} &  CGPR & 0.921 & {\bf 0.927} & {\bf 0.899}  \\
&DPR & 0.820 & 0.575 & 0.002 \\
&GPPI & {\bf 0.995} & 0.916 & 0.654
\end{tabular}
\end{ruledtabular}
\end{table}

\begin{table}[h]
\caption{\label{tb:ASF}%
Comparison of the coefficient of determination ($R^2_I$) for the amplitude sensitivity function (ASF) between the proposed method (CGPR) and an existing method (DPR). 
Methods with the highest $R^2_I$ are highlighted in {\bf bold}. 
}
\begin{ruledtabular}
\begin{tabular}{ccccc}
 $\gamma$ & Method & $\sigma_y=0.01$ & $\sigma_y=0.1$ & $\sigma_y=1.0$ \\ \hline
\multirow{2}{*}{0.2} &  CGPR & {\bf 0.989} & {\bf 0.983} & {\bf 0.976} \\
&DPR& 0.971 & 0.937 & 0.673 \\ \hline
\multirow{2}{*}{0.4} &  CGPR & {\bf 0.994} & {\bf 0.992} & {\bf 0.986}  \\
&DPR & 0.977 & 0.919 & 0.678  \\ \hline
\multirow{2}{*}{0.6} &  CGPR & {\bf 0.977} & {\bf 0.979} & {\bf 0.916} \\
&DPR & 0.953 & 0.878 & 0.651  \\ \hline
\multirow{2}{*}{0.8} &  CGPR & {\bf 0.928} & {\bf 0.939} & {\bf 0.918}  \\
&DPR & 0.906 & 0.836 & 0.615  \\ \hline
\multirow{2}{*}{1.0} &  CGPR & {\bf 0.838} & {\bf 0.875} & {\bf 0.848}  \\
&DPR & 0.821 & 0.763 & 0.559
\end{tabular}
\end{ruledtabular}
\end{table}

In summary, the proposed method provides more accurate and robust inference of dynamical properties than conventional approaches, particularly in the presence of strong observation noise. Moreover, the method remains effective even when the system is far from the bifurcation point, provided that the underlying dynamics are topologically equivalent to the Hopf normal form.

%%%%%%%%%%%%%%%%%%%%%%%%%%%%%%%%%%%%%%%%%%%%
\section{Discussion}\label{sec:discussion}
%%%%%%%%%%%%%%%%%%%%%%%%%%%%%%%%%%%%%%%%%%%%
In this study, we proposed a method for transforming oscillatory time-series data into a representation consistent with the Hopf normal form, thereby enabling the inference of intrinsic dynamical properties such as the natural frequency, Floquet exponent, and asymptotic phase.   
The central idea of the proposed approach is to embed the Hopf normal form within a state-space modeling framework, allowing the transformation from observed data to the latent normal-form representation to be carried out probabilistically. 
Specifically, the mapping from observations to latent states is formulated through Bayesian inference, which provides robustness against observation noise and measurement uncertainty. In addition, we introduced an interpolation scheme for latent-state estimates over unobserved regions of the observation space using complex Gaussian process regression (CGPR), and demonstrated its effectiveness in estimating phase and amplitude sensitivity functions.

Conventional data-driven approaches for identifying oscillatory dynamics are largely based on phase and amplitude reduction techniques, including direct methods \citep{Galan05,Wilson23}, Dynamic Mode Decomposition and its extensions \citep{Schmid10,Williams15,Kutz16}, polynomial regression \citep{Namura22}, Gaussian process regression \citep{Yamamoto25}, autoencoders \citep{Yawata24}, and system-identification methods \citep{Wilson20,Wilson21,Tanaka25}. In contrast to these approaches, the proposed framework provides an alternative perspective that does not explicitly rely on phase--amplitude decomposition.

From the viewpoint of normal-form theory, \citet{Wilson25} recently proposed a method for identifying dynamical systems near Hopf bifurcation by analytically relating the phase and amplitude response curves of the normal form to those estimated directly from data. In a related direction, \citet{Araya26} proposed a data-driven framework for inferring coupled Stuart--Landau equations from waveform measurements. Although the present study shares the objective of fitting Hopf normal forms to observed data, the proposed method differs fundamentally in its probabilistic formulation within a state-space framework. This formulation enables principled treatment of observation noise, accommodates partial observability, and provides uncertainty quantification, which are generally unavailable in deterministic reduction-based approaches. These features make the proposed method particularly suitable for applications involving noisy, incomplete, or transient data.

Although the proposed method successfully recovered intrinsic dynamical properties from noisy time series generated by the van der Pol oscillator, its performance may degrade when applied to highly nonlinear or stiff systems involving multiple time scales. In such settings, identifiability of the latent normal-form coordinates may become challenging, particularly when the observation function is strongly nonlinear or when only partial measurements are available. One possible direction for addressing these limitations is to employ more flexible representations of the observation function in \eqref{eq:observation_model}, replacing the polynomial basis with richer function classes such as adaptive basis expansions \citep{Donner25} or Gaussian process models \citep{Deisenroth09,Turner10}. Such extensions may improve the expressive power of the observation model while preserving the probabilistic structure of the framework.

From a computational perspective, the proposed framework requires sequential inference procedures, including filtering and smoothing, together with CGPR-based interpolation, which may limit scalability for long time series or high-dimensional observations. Developing efficient approximations, sparse representations, or reduced-rank formulations for both the state-space inference and Gaussian process components \cite{Roweis99,quinonero05,Titsias09,Solin14,Sarkka13} will therefore be important for practical large-scale applications.

Furthermore, the current formulation is restricted to systems undergoing a supercritical Hopf bifurcation. Although this setting captures an important class of oscillatory dynamics, the proposed framework naturally suggests extensions to normal forms associated with other types of bifurcations. For example, incorporating saddle-node on invariant circle (SNIC) or pitchfork bifurcations would broaden the applicability of the method to a wider range of dynamical phenomena. In principle, such normal forms can be embedded within the same probabilistic state-space framework.

Finally, although the present study focused on synthetic data, applications to real-world systems introduce additional challenges, including nonstationarity, measurement artifacts, and model mismatch. Addressing these issues, together with extending the framework to a broader class of bifurcations and incorporating more expressive observation models, constitutes an important direction for future research.

%%%%%%%%%%%%%%%%%%%%%%%%%%%%%%%%%%%%%%%%%%
\begin{acknowledgments}
The authors thank Taichi Yamamoto, Norihisa Namura, and Hiroya Nakao for generously providing the codes used in this study. 
This research was partially supported by the Japan Society for the Promotion of Science (JSPS) KAKENHI (Nos. JP22H03695, JP23K24950), and JST FOREST (No. JPMJFR232O) to R.K. 
\end{acknowledgments}
%%%%%%%%%%%%%%%%%%%%%%%%%%%%%%%%%%%%%%%%%%

%%%%%%%%%%%%%%%%%%%%%%%%%%%%%%%%%%%%%%%%%%%%
%%%%%%%%%%%%%%%%%%%%%%%%%%%%%%%%%%%%%%%%%%%%
\appendix
%%%%%%%%%%%%%%%%%%%%%%%%%%%%%%%%%%%%%%%%%%%%
%%%%%%%%%%%%%%%%%%%%%%%%%%%%%%%%%%%%%%%%%%%%
\section{Phase--amplitude representation}
\label{appendix:phase_amplitude}
%%%%%%%%%%%%%%%%%%%%%%%%%%%%%%%%%%%%%%%%%%%%
In this appendix, we summarize the conventional phase--amplitude representation of limit-cycle oscillators \citep{Kuramoto84,Wedgwood13,Nakao16,Shirasaka17,shirasaka2020phase}, and discuss its relation to the Hopf normal form \eqref{eq:normal_form}. We consider a limit-cycle oscillator \eqref{eq:ode_hopf} possessing a stable hyperbolic limit-cycle solution $\bm{y}^{lc}(t)$ with period $T$, satisfying $\bm{y}^{lc}(t+T)=\bm{y}^{lc}(t)$, and denote its basin of attraction by $A \subseteq \mathbb{R}^D$.

To characterize the oscillator state $\bm{y}$, we introduce a phase function $\Theta(\bm{y}) : A \to [0,2\pi)$ that assigns the asymptotic phase to each state in the basin of attraction. First, the phase for a state $\bm{y}^{lc}(t)$ on the limit cycle is defined as
\begin{equation}
\Theta(\bm{y}^{lc}(t)) = \omega t \pmod{2\pi},
\end{equation}
where $\omega = 2\pi/T$ denotes the natural frequency. By definition, the phase increases uniformly along the limit cycle according to
\begin{equation}
\frac{d}{dt}\Theta(\bm{y}^{lc}(t)) = \omega.
\label{eq:phasefunc_lm}
\end{equation}
The phase origin, corresponding to phase $0$, is given by $\bm{y}^{lc}(0)$ and can be chosen arbitrarily. A common convention is to define the phase origin as the point on the periodic orbit at which the first coordinate attains its maximum value. In the following, we denote the state on the limit cycle with phase $\theta$ by $\bm{y}^{lm}(\theta)$. The phase function can be uniquely extended over the entire basin of attraction $A$ so that \eqref{eq:phasefunc_lm} holds for any trajectory $\bm{y}(t)$ in $A$:
\begin{equation}
\frac{d}{dt}\Theta(\bm{y}(t)) = \omega.
\label{eq:phasefunc}
\end{equation}

Next, we introduce the amplitude function. Analogously to the phase function, we define a scalar amplitude function $R : A \to \mathbb{R}$ satisfying
\begin{equation}
\frac{d}{dt} R(\bm{y}(t)) = \lambda R(\bm{y}(t)),
\label{eq:amplitudefunc}
\end{equation}
where $\lambda < 0$ denotes the Floquet exponent of the limit cycle with the largest nonzero real part. Since trajectories converge asymptotically to the limit cycle, the amplitude decays exponentially and satisfies $R(\bm{y}^{lm}(\theta)) = 0$ for all points on the limit cycle.

The phase sensitivity function (PSF) is defined as the gradient of the phase function evaluated along the limit cycle:
\begin{equation}
\bm{Z}(\theta)
=
\nabla \Theta(\bm{y}) \big|_{\bm{y}=\bm{y}^{lm}(\theta)},
\end{equation}
which characterizes the linear response of the oscillator phase to weak perturbations applied at phase $\theta$. Similarly, the amplitude sensitivity function (ASF) is defined as the gradient of the amplitude function along the limit cycle:
\begin{equation}
\bm{I}(\theta)
=
\nabla R(\bm{y}) \big|_{\bm{y}=\bm{y}^{lm}(\theta)},
\label{eq:ASF_def}
\end{equation}
which characterizes the linear response of the oscillator amplitude to weak perturbations applied at phase $\theta$. When the governing equations of the oscillator are known, the PSF and ASF can be computed numerically using adjoint-based methods \citep{Takata21,Namura22}.

The relationship between the phase--amplitude representation and the Hopf normal form becomes explicit when the Hopf normal form is expressed in polar coordinates. By comparing \eqref{eq:hopf_w} with \eqref{eq:phasefunc}, we observe that the polar angle of the Hopf normal form corresponds to the asymptotic phase, up to an arbitrary choice of phase origin. Furthermore, the amplitude equation \eqref{eq:hopf_r} agrees with \eqref{eq:amplitudefunc} in the vicinity of the limit cycle, which can be verified by redefining the amplitude variable of the Hopf normal form as $R(t)=1-r(t)$. Consequently, the ASF derived from the Hopf normal form is consistent with the conventional definition given in \eqref{eq:ASF_def}.

%%%%%%%%%%%%%%%%%%%%%%%%%%%%%%%%%%%%%%%%%%%%
\section{Filtering and Smoothing Procedure}\label{appendix:filtersmoothing}
%%%%%%%%%%%%%%%%%%%%%%%%%%%%%%%%%%%%%%%%%%%%

In this appendix, we present the means and covariances of the prediction, filtering, and smoothing densities obtained via moment matching. 
For notational simplicity, we rewrite \eqref{eq:state_model} and \eqref{eq:observation_model} as
\begin{align}
\bm{x}_{t+1} &= G\bm{x}_t + \bm{g}(\bm{x}_t) + \bm{\xi}_t, \\
\bm{y}_t &= C\bm{z}_t + \bm{\zeta}_t,
\end{align}
where
\begin{align}
G = 
\begin{pmatrix}
1+\alpha\Delta & -\omega\Delta \\ 
\omega\Delta & 1+\alpha\Delta
\end{pmatrix}, 
\quad
\bm{g}(\bm{x}) = - \alpha\Delta \lVert \bm{x} \rVert^2 \bm{x},
\end{align}
\begin{align}
C = (C_0,C_1,\ldots,C_p)
\in \mathbb{R}^{2 \times (p+1)(p+2)/2},
\label{eq:coef_poly}
\end{align}
and
\begin{align}
\bm{z}_t =
\left(1,\bm{x}_t,{\bm{x}_t^{(1)}}^\mathsf{T},\ldots,{\bm{x}_t^{(p)}}^\mathsf{T}\right)^\mathsf{T}
\in \mathbb{R}^{(p+1)(p+2)/2}
\end{align}
denotes the concatenated state vector.

%%%%%%%%%%%%%%%%%%%%
\subsection{Filtering}
%%%%%%%%%%%%%%%%%%%%
The mean and covariance of the prediction density \eqref{eq:prediction} are given by
\begin{align}
\bm{\mu}^p_t 
&= \mathbb{E}_{\Theta}[\bm{x}_t \mid \bm{y}_{1:t-1}] \nonumber\\
&= G\bm{\mu}^f_{t-1} + \mathbb{E}_{\Theta}[\bm{g}(\bm{x}_{t-1}) \mid \bm{y}_{1:t-1}],
\end{align}
and
\begin{widetext}
\begin{align}
\Sigma^p_{t|t-1} 
&= \mathbb{E}_{\Theta}\big[(\bm{x}_{t}-\bm{\mu}^p_t)(\bm{x}_{t}-\bm{\mu}^p_t)^\mathsf{T} \mid \bm{y}_{1:t-1}\big] \nonumber\\
&= \mathbb{E}_{\Theta}\big[(G\bm{x}_{t-1} + \bm{g}(\bm{x}_{t-1}))
(G\bm{x}_{t-1} + \bm{g}(\bm{x}_{t-1}))^\mathsf{T} \mid \bm{y}_{1:t-1}\big] 
+ \Sigma_x - \bm{\mu}^p_t(\bm{\mu}^p_t)^\mathsf{T},
\end{align}
\end{widetext}
where $\mathbb{E}_{\Theta}[\cdot \mid \bm{y}_{1:t-1}]$ denotes expectation with respect to 
$\mathcal{N}(\bm{\mu}^f_{t-1},\Sigma^f_{t-1})$.

To compute the mean and covariance of the filtering density \eqref{eq:filtering}, 
we approximate the joint density $p(\bm{x}_t,\bm{y}_t \mid \bm{y}_{1:t-1},\Theta)= p(\bm{y}_t \mid \bm{x}_t,\Theta)\, p(\bm{x}_t \mid \bm{y}_{1:t-1},\Theta)$ by a Gaussian. 
Under this approximation, the conditional density $p(\bm{x}_t \mid \bm{y}_{1:t},\Theta)$ is Gaussian, whose mean and covariance are given by
\begin{widetext}
\begin{align}
\bm{\mu}^f_t
&= \bm{\mu}^p_t + \mathrm{cov}_{\Theta}(\bm{x}_t,\bm{y}_t \mid \bm{y}_{1:t-1})
\mathrm{var}_{\Theta}(\bm{y}_t \mid \bm{y}_{1:t-1})^{-1}
(\bm{y}_t- \bm{y}^p_t), \\
\Sigma^f_t
&= \Sigma^p_t - \mathrm{cov}_{\Theta}(\bm{x}_t,\bm{y}_t \mid \bm{y}_{1:t-1})
\mathrm{var}_{\Theta}(\bm{y}_t \mid \bm{y}_{1:t-1})^{-1}
\mathrm{cov}_{\Theta}(\bm{x}_t,\bm{y}_t \mid \bm{y}_{1:t-1})^\mathsf{T}.
\end{align}
\end{widetext}
Here, $\bm{y}^p_t$, $\mathrm{var}_{\Theta}(\bm{y}_t \mid \bm{y}_{1:t-1})$, and 
$\mathrm{cov}_{\Theta}(\bm{x}_t,\bm{y}_t \mid \bm{y}_{1:t-1})$ are given by
\begin{align}
\bm{y}^p_t
= \mathbb{E}_{\Theta}[\bm{y}_t \mid \bm{y}_{1:t-1}]
= C\,\mathbb{E}_{\Theta}[\bm{z}_t \mid \bm{y}_{1:t-1}],
\end{align}
\begin{eqnarray}
\mathrm{var}_{\Theta}(\bm{y}_t \mid \bm{y}_{1:t-1})
=&&
C\,\mathbb{E}_{\Theta}\left[ \bm{z}_t\bm{z}_t^\mathsf{T}\mid \bm{y}_{1:t-1} \right]C^\mathsf{T} \nonumber\\
&&+ \Sigma_y - \bm{y}^p_t(\bm{y}^p_t)^\mathsf{T},
\end{eqnarray}
\begin{align}
\mathrm{cov}_{\Theta}(\bm{x}_t,\bm{y}_t \mid \bm{y}_{1:t-1})
&=
\mathbb{E}_{\Theta}\left[ \bm{x}_t\bm{z}_t^\mathsf{T} \mid \bm{y}_{1:t-1} \right] C^\mathsf{T} 
- \bm{\mu}^p_t(\bm{y}^p_t)^\mathsf{T},
\end{align}
where the expectations are taken with respect to 
$\mathcal{N}(\bm{\mu}^p_t,\Sigma^p_t)$.

%%%%%%%%%%%%%%%%%%%%
\subsection{Smoothing}
%%%%%%%%%%%%%%%%%%%%
To derive the smoothing distribution \eqref{eq:smoothing}, we approximate the joint density 
$p(\bm{x}_t,\bm{x}_{t+1} \mid \bm{y}_{1:t},\Theta)$ by a Gaussian via moment matching:
\begin{equation}
p(\bm{x}_t,\bm{x}_{t+1} \mid \bm{y}_{1:t},\Theta) \approx
\mathcal{N}\left(
\begin{pmatrix}
\bm{\mu}^f_t \\ 
\bm{\mu}^p_{t+1} 
\end{pmatrix}, 
\begin{pmatrix}
\Sigma^f_{t} & \Sigma^p_{t,t+1} \\ 
(\Sigma^p_{t,t+1})^\mathsf{T} & \Sigma^p_{t+1}
\end{pmatrix}
\right),
\end{equation}
where $\Sigma^p_{t,t+1} = \mathrm{cov}_{\Theta}(\bm{x}_t,\bm{x}_{t+1} \mid \bm{y}_{1:t})$. 
The conditional density $p(\bm{x}_t \mid \bm{x}_{t+1}, \bm{y}_{1:t},\Theta)$ is then Gaussian with
\begin{align}
\tilde{\bm{\mu}}_t
&= \bm{\mu}^f_t + \Sigma^p_{t,t+1}(\Sigma^p_{t+1})^{-1}(\bm{x}_{t+1}-\bm{\mu}^p_{t+1}), \\
\tilde{\Sigma}_t
&= \Sigma^f_{t} - \Sigma^p_{t,t+1}(\Sigma^p_{t+1})^{-1}(\Sigma^p_{t,t+1})^\mathsf{T}.
\end{align}
Combining this with $p(\bm{x}_{t+1} \mid \bm{y}_{1:T},\Theta)$ yields the smoothing recursion
\begin{align}
p(\bm{x}_t, \bm{x}_{t+1}  \mid  \bm{y}_{1:T},\Theta) 
&= 
p(\bm{x}_t \mid \bm{x}_{t+1}, \bm{y}_{1:t},\Theta)p(\bm{x}_{t+1} \mid \bm{y}_{1:T},\Theta) \nonumber\\
&\approx
\mathcal{N}(\tilde{\bm{\mu}}_t,\tilde{\Sigma}_t)
\mathcal{N}(\bm{\mu}^s_{t+1},\Sigma^s_{t+1}) \nonumber\\
&=
\mathcal{N}\left(
\begin{pmatrix}
\bm{\mu}^s_t \\ 
\bm{\mu}^s_{t+1} 
\end{pmatrix}, 
\begin{pmatrix}
\Sigma^s_{t} & \Sigma^s_{t,t+1} \\ 
(\Sigma^s_{t,t+1})^\mathsf{T} & \Sigma^s_{t+1}
\end{pmatrix}
\right),
\end{align}
where
\begin{align}
\bm{\mu}^s_t &= \bm{\mu}^f_t + K_t(\bm{\mu}^s_{t+1}-\bm{\mu}^p_{t+1}), \\
\Sigma^s_t &= \Sigma^f_t + K_t(\Sigma^s_{t+1}-\Sigma^p_{t+1})K_t^\mathsf{T}, \\
\Sigma^s_{t,t+1} &= K_t\Sigma^s_{t+1}, \\
K_t &= \Sigma^p_{t,t+1}(\Sigma^p_{t+1})^{-1}.
\end{align}

%%%%%%%%%%%%%%%%%%%%%%%%%%%%%%%%%%%%%%%%%%%%
\section{EM Algorithm}\label{appendix:em}
%%%%%%%%%%%%%%%%%%%%%%%%%%%%%%%%%%%%%%%%%%%%
The parameter update equations are obtained by maximizing the auxiliary function \eqref{eq:qfunc} with respect to each parameter. 
The resulting updates are given by
\begin{widetext}
\begin{align}
\hat{\alpha}
&= 
\frac{1}{\Delta}
\frac{
\sum_{t=1}^T\mathbb{E}_{\Theta}\!\left[ (1-\lVert\bm{x}_{t-1}\rVert^2)\bm{x}_t^\mathsf{T}\bm{x}_{t-1} \mid \bm{y}_{1:T}\right]
-
\sum_{t=1}^T \mathbb{E}_{\Theta}\!\left[ (1-\lVert\bm{x}_{t-1}\rVert^2)\bm{x}_{t-1}^\mathsf{T}\bm{x}_{t-1} \mid \bm{y}_{1:T}\right]
}{
\sum_{t=1}^T \mathbb{E}_{\Theta}\!\left[ (1-\lVert\bm{x}_{t-1}\rVert^2)^2\bm{x}_{t-1}^\mathsf{T}\bm{x}_{t-1} \mid \bm{y}_{1:T}\right]
}, \\
\hat{\omega} &= 
\frac{1}{\Delta}
\frac{
\sum_{t=1}^T\mathbb{E}_{\Theta}[\bm{x}_t^\mathsf{T} J \bm{x}_{t-1} \mid \bm{y}_{1:T}]
}{
\sum_{t=1}^T\mathbb{E}_{\Theta}[ \bm{x}_{t-1}^\mathsf{T}\bm{x}_{t-1} \mid \bm{y}_{1:T}]
}, 
\quad 
J = 
\begin{pmatrix}
0 & -1 \\
1 & 0
\end{pmatrix},
\\
\hat{\Sigma}_x
&= \frac{1}{T}
\sum_{t=1}^T \mathbb{E}_{\Theta}\!\left[
(\bm{x}_t-G\bm{x}_{t-1}-\bm{g}(\bm{x}_{t-1}))
(\cdot)^\mathsf{T} \mid \bm{y}_{1:T}\right], \\
\hat{C} &= 
\left( \sum_{t=1}^T\bm{y}_t\mathbb{E}_{\Theta}[\bm{z}_t\mid \bm{y}_{1:T}]^\mathsf{T} \right)
\left( \sum_{t=1}^T \mathbb{E}_{\Theta}[ \bm{z}_t\bm{z}_t^\mathsf{T}\mid \bm{y}_{1:T}] \right)^{-1}, \\
\hat{\Sigma}_y &= 
\frac{1}{T}\sum_{t=1}^T
\mathbb{E}_{\Theta}\!\left[ (\bm{y}_t - \bm{h}(\bm{x}_t))(\bm{y}_t - \bm{h}(\bm{x}_t))^\mathsf{T} 
\mid \bm{y}_{1:T} \right], \\
\hat{\bm{\mu}}_0 &= \bm{\mu}^s_0, 
\quad 
\hat{\Sigma}_0 = \Sigma^s_0.
\end{align}
\end{widetext}

\bibliography{references}

\end{document}